\documentclass[aps,twocolumn,superscriptaddress, nofootinbib, longbibliography, notitlepage]{revtex4-1}  

\usepackage{amssymb,amsmath,amsfonts,amsbsy}
\usepackage{color}
\usepackage{enumerate}
\usepackage{graphicx}
\usepackage{pifont}

\usepackage{subfigure}

\usepackage{cancel}

\RequirePackage[colorlinks,citecolor=blue,urlcolor=blue,linkcolor=blue]{hyperref}

\usepackage[normalem]{ulem}

\begin{document}

\title{Constraining gravitational wave propagation using pulsar timing array correlations}

\author{Reginald Christian Bernardo}
\affiliation{Institute of Physics, Academia Sinica, Taipei 11529, Taiwan}
\email{rbernardo@gate.sinica.edu.tw}

\author{Kin-Wang Ng}
\affiliation{Institute of Physics, Academia Sinica, Taipei 11529, Taiwan}
\affiliation{Institute of Astronomy and Astrophysics, Academia Sinica, Taipei 11529, Taiwan}
\email{nkw@phys.sinica.edu.tw}

\begin{abstract}
Pulsar timing arrays (PTA) are a promising probe to the cosmologically novel nanohertz gravitational wave (GW) regime through the stochastic GW background. In this work, we consider subluminal GW modes as a possible source of correlations in a PTA, utilizing the public code PTAfast and the 12.5 years correlations data by NANOGrav, which we hypothesize are sourced by GWs. Our results show no evidence in support of tensor-- or vector--induced GW correlations in the data, and that vector correlations are disfavored. This places an upper bound to the graviton mass, $m_{\rm g} \lesssim 10^{-22}$ eV, characteristic of the PTA GW energy scale.
\end{abstract}

\maketitle

\section{Introduction}
\label{sec:introduction}

The next gravitational wave (GW) astronomy breakthrough is the detection of the stochastic gravitational wave background (SGWB) -- a galaxy size superposition of GWs from various cosmological sources -- that correlates astronomical observations \cite{Burke-Spolaor:2018bvk, NANOGrav:2020spf}. Along this theme, it was advocated decades ago that the precise timing of the emission of radio pulses by nearby millisecond pulsars is equivalent to a galactic size GW detector \cite{Detweiler:1979wn}. This proposal reached fruition in so--called `pulsar timing array' (PTA) missions that aim to capture the wavelike features of the SGWB, and work as probes of the cosmologically novel nanohertz GW regime \cite{Romano:2016dpx, Romano:2019yrj}.

As the CMB is host to physics during last scattering, the SGWB is host to a variety of sources tied to the early Universe, such as supermassive black hole binary coalescences \cite{Phinney:2001di, Wyithe:2002ep, Shannon:2015ect}, and more exotic phenomena like cosmic strings \cite{Ellis:2020ena, Buchmuller:2021mbb, Hindmarsh:2022awe}, phase transitions \cite{NANOGrav:2021flc, Xue:2021gyq}, and primordial black holes \cite{Nakama:2016gzw, Chen:2019xse}, among others \cite{Caprini:2018mtu, Sharma:2021rot}. PTA astronomy caters to such exciting physics that come with the SGWB.

The present PTAs, spearheaded by the North American Nanohertz Observatory for Gravitational Waves (NANOGrav) \cite{NANOGrav:2020bcs}, European PTA \cite{Chen:2021rqp}, and Parkes PTA \cite{Goncharov:2021oub}, working together under the banner International PTA (IPTA) \cite{2010CQGra..27h4013H, Chen:2021ncc}, finds strong evidence for a stochastic common process across millisecond pulsars. However, the present data have yet to reach maturity to be able to tell with confidence whether this common process is due to the SGWB. This is where PTA cross correlations play a significant role, since the SGWB gives a distinct quadrupolar spatial correlation \cite{Hellings:1983fr, Ng:2021waj, Allen:2022dzg, Allen:2022ksj} that cannot be associated with systematic effects such as clock error and ephemeris uncertainty \cite{2016MNRAS.455.4339T, Roebber:2019gha, NANOGrav:2020tig}. As the PTAs continue to monitor more pulsars, SGWB correlations should soon manifest in the data \cite{Pol:2022sjn}, marking PTA science precision era. 

Adding to the science prospects of PTAs, in this work, we take a conservative stand on SGWB correlations and consider subluminal GW modes as a source of the observed cross correlations in a PTA \cite{Qin:2020hfy, Bernardo:2022rif, Bernardo:2022xzl}. We use the public code PTAfast \cite{2022ascl.soft11001B} and the correlations data set by NANOGrav to constrain the GW speed, and henceforth give its take on the graviton mass. 

The rest of this paper proceeds as follows. We first address the feasibility of subluminal GWs given the known tight constraint on GW propagation at sub-kilohertz frequencies \cite{LIGOScientific:2017vwq} (Section \ref{sec:subluminal_gws}). We then briefly go over the power spectrum method that is implemented to calculate the SGWB spatial correlation signals (Section \ref{subsec:sgwb_correlations}), and describe the data (Section \ref{subsec:data}) to be considered in upcoming data analysis (Section \ref{subsec:bayesian_analysis}). We present our main results in two parts, beginning with only subluminal GW modes (Section \ref{subsec:graviton_mass}), and following it up with an assessment of its significance (Section \ref{subsec:subluminal_hd_mon}). We conclude with a summary of our results and draw future directions on subluminal GW propagation (Section \ref{sec:conclusions}).

\section{Subluminal GWs}
\label{sec:subluminal_gws}

Before we push through, we discuss the feasibility of subluminal tensors in the PTA/nanohertz GW band, particularly in light of the GW astronomy constraint $| 1 - v | \lesssim 10^{-15}$ at $f \sim 100$ hertz \cite{LIGOScientific:2017vwq, LIGOScientific:2017zic}.

We start by writing down the dispersion, $\omega^2 = k^2 + m_{\rm g}^2$, expressed in terms of the velocity, $v = d\omega/dk$, wave number $k$, and particle (graviton) mass $m_{\rm g}$, as
\begin{equation}
    1 - v = 1 - \sqrt{ 1 - \left( \dfrac{m_{\rm g}}{\omega} \right)^2 } \,.
\end{equation}
Thus, providing $1 - v \sim 10^{-15}$, and taking physical considerations into account, $v < 1$ and $m_{\rm g} \leq \omega$, we find
\begin{equation}
    \left( \dfrac{m_{\rm g}}{\omega} \right)^2 \lesssim 2 \times 10^{-15} \,.
\end{equation}
At $f \sim 100$ hertz (or $\omega \sim 2 \pi 100$ hertz), we see that the LIGO/Virgo band, in this crude calculation, translates to a graviton mass upper bound of $m_{\rm g} \lesssim 10^{-19}$ eV. In general, the GW speed constraint converts to
\begin{equation}
    m_{\rm g} \lesssim \left( \dfrac{f}{100 \ {\rm hertz}} \right) \times 10^{-19} \ {\rm eV} \,.
\end{equation}
This expression shows that in PTAs (at much lower frequencies, $f \sim 1$ per year) the graviton mass bounds are always going to be acceptable within the LIGO/Virgo GW speed constraint. In other words, subluminal GW propagation is allowed by the present constraints \cite{deRham:2018red}.

Keeping this in mind, we proceed to the SGWB correlations induced by subluminal GWs in PTA.

\section{PTA Correlations}
\label{sec:pta_correlations}

We briefly describe the PTA correlations in terms of the power spectrum, and the data which we will use. But, first we discuss the feasibility of subluminal tensors in the nanohertz GW band.

\subsection{SGWB correlations}
\label{subsec:sgwb_correlations}

The SGWB spatially correlates the time of arrival perturbations of the radio pulses emitted by nearby millisecond pulsars. This \textit{angular} correlation, $\gamma_{ab}\left(\zeta\right)$, can be written in terms of the power spectrum, $C_l$'s, as
\begin{equation}
    \gamma_{ab}\left(\zeta\right) = \sum_{l} \dfrac{2l+1}{4\pi} C_l P_l\left(\cos \zeta\right) \,,
\end{equation}
where $\zeta$ is the angular separation of pulsars $a$ and $b$ in the sky.

For this work, we consider only tensor and vector polarizations. Subluminal scalar GWs were discussed in \cite{Bernardo:2022vlj}. In general, the power spectrum multipoles can be written as
\begin{equation}
    C_l = \dfrac{ J_l \left( fD_a \right) J_l^* \left( fD_b \right) }{\sqrt{\pi}} \,,
\end{equation}
where for tensor modes the $J_l(y)$'s are
\begin{equation}
    J_l\left( y \right) = \sqrt{2} \pi i^l \sqrt{ \dfrac{(l + 2)!}{(l - 2)!} } \int_0^{2\pi y v} \dfrac{dx}{v} e^{ix/v} \dfrac{j_l\left( x \right)}{x^2} \,,
\end{equation}
whereas for vector modes the $J_l(y)$'s are instead
\begin{equation}
    J_l\left( y \right) = 2 \sqrt{2} \pi i^l \sqrt{l\left(l+1\right)} \int_0^{2\pi y v} \dfrac{dx}{v} e^{ix/v} \dfrac{d}{dx} \left( \dfrac{j_l(x)}{x} \right) \,.
\end{equation}
Above, $j_l(x)$'s are the spherical Bessel functions, $f$ is the GW frequency, $v$ is the GW speed, and $D_i$'s are the distances to the pulsars.

The overlap reduction function (ORF), $\Gamma_{ab}\left(\zeta\right)$, or the PTA's spatial correlation observable, is obtained simply by normalizing $\gamma_{ab}\left(\zeta\right)$ such that $\Gamma_{ab}\left(0\right) = 0.5$ for tensors with $v = 1$. We refer the reader to \cite{Bernardo:2022rif, Bernardo:2022xzl} for a detailed description of this procedure, and a derivation of the results put forward in this section.

The autocorrelation, $\gamma_{aa}$, embodying a pulsar's correlation with itself, is also of relevance in this work. This physical quantity admits analytical expressions, as one dimensional integrals, which we give below for the tensor and vector modes \cite{Bernardo:2022rif}: for tensor polarizations the autocorrelation is
\begin{equation}
\label{eq:gammaaa}
    \gamma_{aa} = \int_0^\pi \dfrac{d\theta}{\sqrt{4\pi}} \left( \frac{2 \pi  \sin ^5 \theta  \sin ^2(\pi  fD (1 + v \cos \theta))}{(1 + v \cos \theta )^2} \right) \,,
\end{equation}
and for vector polarizations the expression is
\begin{equation}
\label{eq:gammaaa_V}
    \gamma_{aa} = \int_0^\pi \dfrac{d\theta}{\sqrt{4\pi}} \left( \frac{ 8\pi \sin^3 \theta \cos^2 \theta \sin ^2(\pi  fD (1 + v \cos \theta))}{(1 + v \cos \theta )^2} \right) \,.
\end{equation}

For our purposes, we apply PTAfast \cite{2022ascl.soft11001B} which implements all the intervening calculations, and provides the ORF's and autocorrelations relevant for comparison with observations.

\subsection{Data}
\label{subsec:data}

The NANOGrav \cite{NANOGrav:2020bcs}, among other PTAs \cite{Chen:2021rqp, Goncharov:2021oub}, confirm the existence of a stochastic common spectrum process. This is not sufficient to tell unequivocally whether the SGWB is out there, but it nonetheless is able to give bounds to the SGWB amplitude. In particular, for a spectral density, characteristic of supermassive black hole binaries, the NANOGrav gives the GW characteristic strain, $A_{\rm CP}$, a median $1.92 \times 10^{-15}$, and $5-95\%$ quantiles $1.37-2.67 \times 10^{-15}$ at $f = 1$ yr$^{-1}$.

Spatial correlations, on the other hand, reveal the SGWB, as it perturbs the arrival of the radio pulses from the millisecond pulsars, delaying some while advancing others, depending on when and where they stand on the wave. However, quite similar with the CMB, digging up this GW correlation from the noise requires the precise monitoring of as many as possible pulsars for long periods. For the moment, the PTAs altogether keep an eye on fewer than a hundred millisecond pulsars, which remain unperceptive of the SGWB, but nonetheless present nontrivial features in the correlation.

For this work, we make use of the correlation data points by NANOGrav \cite{NANOGrav:2020bcs} that measures the average cross correlated power (CCP) across the pulsars in the PTA, ${\rm CCP}\left(\zeta\right) \pm \Delta {\rm CCP}\left(\zeta\right)$. In addition, we utilize the GW amplitude of the common process as a data point, $A_{\rm CP}^2 = A^2 \Gamma_{aa}$, that we associate with the autocorrelation.

The 12.5 years noise--marginalized cross correlation data by NANOGrav comprises of timing residual observations of 45 millisecond pulsars, which make up 990 pulsar pairs that we distribute evenly over 15 bins in angular space \cite{NANOGrav:2020bcs}. The error $\Delta {\rm CCP}$ thus contains all statistical and systematic uncertainty leftovers after marginalizing over the intrinsic and nuisance parameters pertaining to each millisecond pulsar.

\section{Results}
\label{sec:results}

We present our main results in this section, starting with our data analysis implementation (Section \ref{subsec:bayesian_analysis}), and then moving on to the correlations (Sections \ref{subsec:graviton_mass} and \ref{subsec:subluminal_hd_mon}).

\subsection{Bayesian analysis}
\label{subsec:bayesian_analysis}

We model the CCP in a PTA as $A^2 \Gamma_{ab}\left(\zeta\right)$, making up a two dimensional parameter space, $v \times A^2$, of the GW speed $v$ and the amplitude squared $A^2$. We compare this with the data through the likelihood function $\mathcal{L}$, which we take to be
\begin{equation}
\label{eq:loglike}
    \log {\cal L} = -\dfrac{1}{2}\sum_{\zeta} \left(\dfrac{ A^2
 \Gamma_{ab}\left(\zeta\right) - {\rm CCP}\left(\zeta\right) }{ 
\Delta {\rm CCP}\left(\zeta\right) }\right)^2 \,,
\end{equation}
where the sum runs over the angular separations of pulsar pairs in the PTA.

We also make use of the amplitude of the common process, which we symmetrize for our purposes to be $A_{\rm CP}^2 = \left( 3.68 \pm 1.58 \right) \times 10^{-30}$. This conservative value is obtained from the median and larger error bar of the asymmetric bound of the common process \cite{NANOGrav:2020bcs}. Admittedly, this is not the most faithful reconstruction from the best estimate, but it encompasses all the relevant values. Another point that we bring up once more, as alluded to previously, is that this value $A_{\rm CP}$ was obtained assuming that the SGWB comes mainly from supermassive black hole binaries, and so carries with it a mild model dependence despite the large uncertainty. We use it anyway for the analysis, but guarantee that the main results of this paper holds even with only the correlations, which dominate the bulk of the data.

Now that the model, data, and the likelihood are laid out, we obtain the posterior, $P(\theta|D)$, through Bayes theorem, $P\left(\theta|D\right) \propto P\left(\theta\right) {\cal L}$, by sampling over the parameter space $\theta =\left(v , A^2\right)$ via a Markov chain Monte Carlo (MCMC) algorithm \cite{Trotta:2008qt}. We consider flat priors $P\left(\theta\right)$ for both the parameters, $v \in [0.01, 1.00]$ and $A^2 \in [0.01, 50] \times 10^{-30}$, and as a starting point (each time) take the references $v \in (0.48, 0.52)$ and $A^2 \in (1, 2) \times 10^{-30}$. The references give a better guarantee when the run has converged since the initial points are always different. This is in addition to specifying a strict Gelman-Rubin convergence criterion of $R - 1 = 10^{-3}$ to mark the end of the sampling.

We implement these steps using PTAfast \cite{2022ascl.soft11001B} together with the cosmology community code Cobaya \cite{Lewis:2019xzd}, and analyze the results using GetDist \cite{Torrado:2020dgo}. Our codes and resulting MCMC chains are publicly accessible in \href{https://github.com/reggiebernardo/PTAfast}{GitHub}.

\subsection{Subluminal GWs and the graviton}
\label{subsec:graviton_mass}

\begin{figure}[h!]
    \centering
    \includegraphics[width = 0.49 \textwidth]{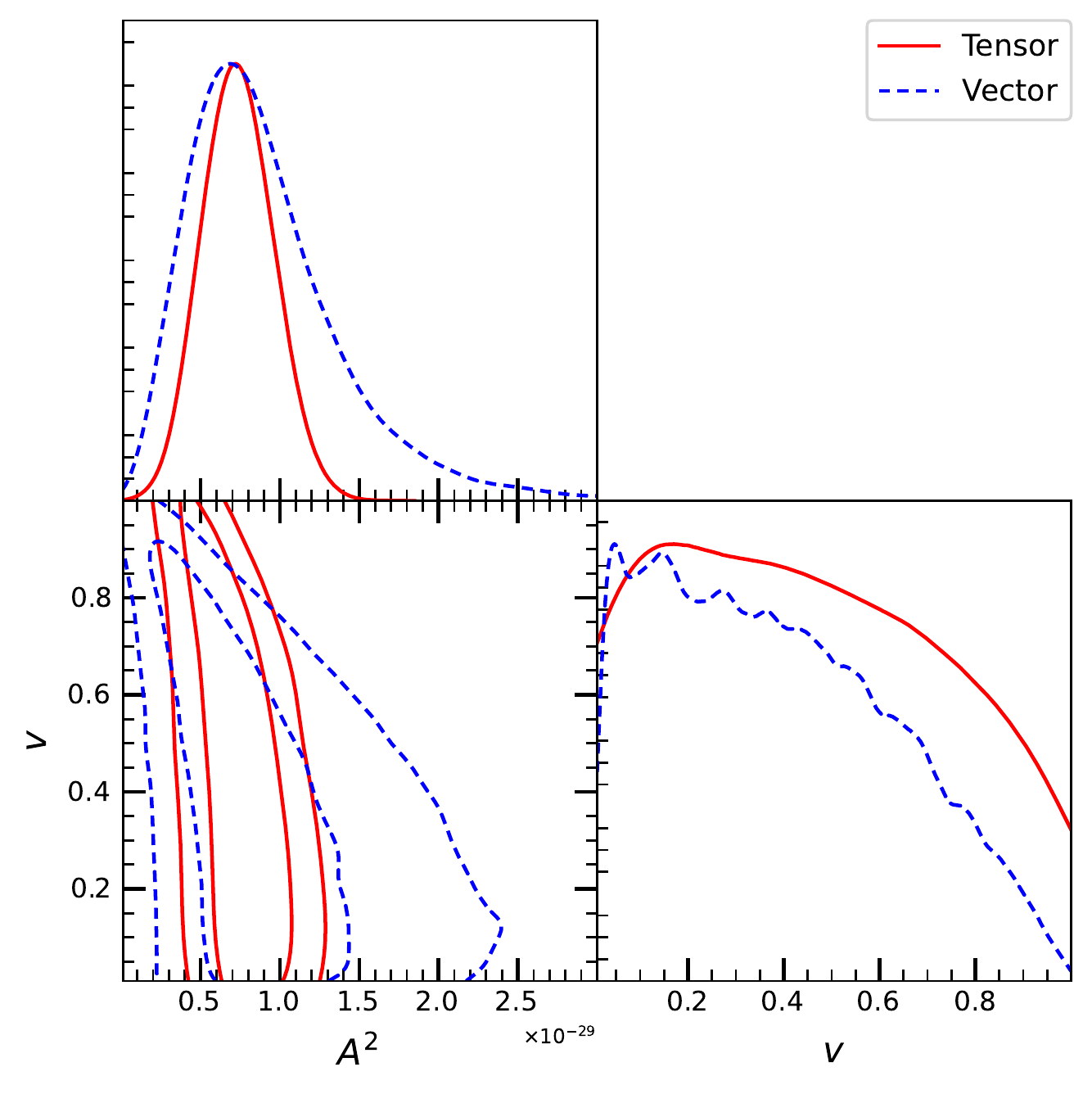}
    \caption{Sampled posterior of the GW speed, $v$, and the amplitude-squared, $A^2$, of subluminal GW modes determined by PTA data \cite{NANOGrav:2020bcs}. Two dimensional contour shows the 68\% and 95\% confidence intervals, respectively, for tensor (red solid) and vector (blue dashed) GW correlations.}
    \label{fig:triangle}
\end{figure}

Figure \ref{fig:triangle} shows the posterior resulting from PTA correlation observations.

The results show that the PTA correlations constrain the GW amplitude and the GW speed, despite the low level of precision of the present correlation measurements. Furthermore, an interesting angle here is that the data supports subluminal GW modes ($v < 1$) more than luminal ones ($v = 1$) at 95\% confidence. We remind that subluminal GW propagation is consistent with the GW speed constraint by LIGO/Virgo \cite{LIGOScientific:2017vwq}. Caution must however be taken care of when interpreting the nonrelativistic speed limit ($v \ll 1$) since this regime is sensitive to the pulsar distances. Granted, the large angle correlations that the current PTAs are able to see are very weakly influenced by how far the pulsars are from the observer \cite{Ng:2021waj}. However, this picture changes when entering the nonrelativistic limit ($v \ll 1$) where the SGWB power spectrum exhibits modes, beyond the quadrupole, that manifest at large angles \cite{Bernardo:2022rif, Bernardo:2022xzl}. Nonetheless, the takeaway of this result is clear: that the uncertainty in the present PTA data is able to support subluminal GW propagation.

We add that these hold regardless of the nature, that is, tensor or vector, of the GWs making up the SGWB. Scalar field induced GWs were considered in \cite{Bernardo:2022vlj} where similar conclusions were drawn. At the same time, there are noticeable differences that discriminate between these tensor and vector modes. For one, the GW amplitude is more constrained for tensors than for vectors, i.e., tensors have narrower $A^2$ posterior compared to vectors. It is also worth adding that the correlations strongly disfavor luminal vector GW modes, which we understand is a reflection of the divergence of vector correlations at luminal speed \cite{NANOGrav:2021ini}. Tensor SGWB correlations are in fact the only ones that does not lose predictability at luminal speeds \cite{Bernardo:2022rif, Bernardo:2022xzl}.

\begin{figure}[h!]
\centering
	\subfigure[ \ GW speed $v$ vs graviton mass $m_{\rm g}$ ]{
		\includegraphics[width = 0.475 \textwidth]{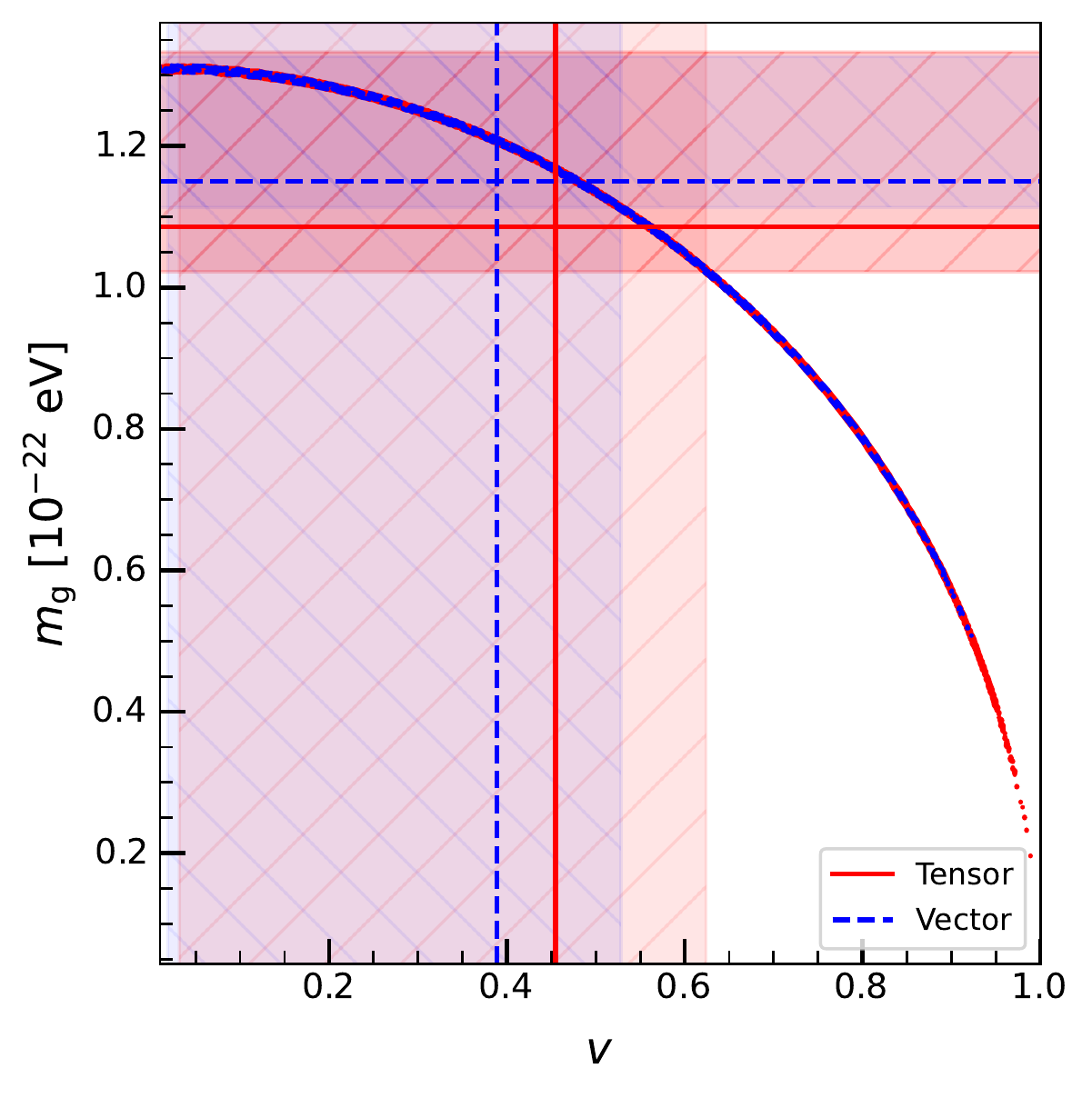}
		}
	\subfigure[ \ graviton mass $m_{\rm g}$ normalized posterior ]{
		\includegraphics[width = 0.45 \textwidth]{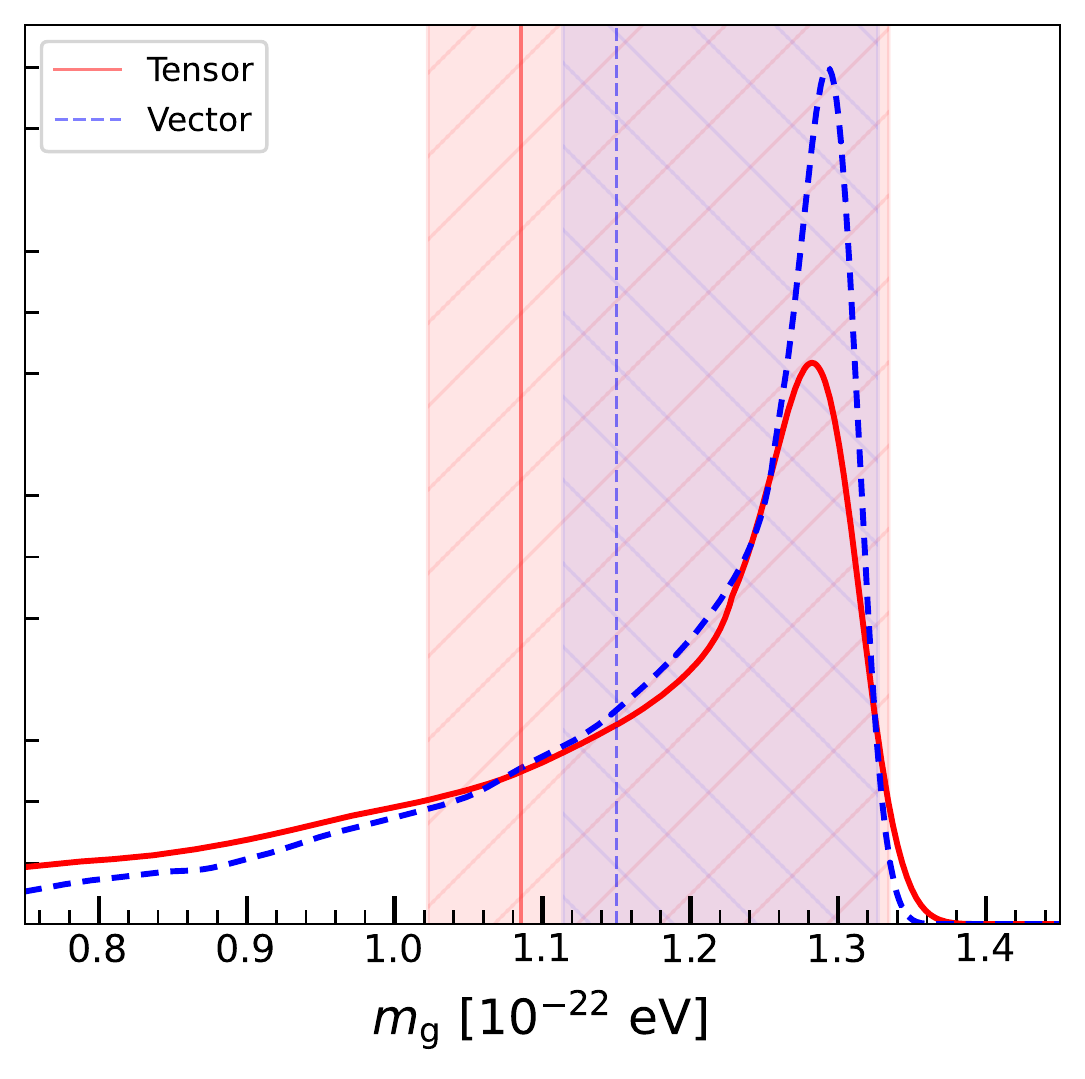}
		}
\caption{Sampled graviton mass posteriors at $f \sim 1$ per year: (a) Two dimensional posterior showing the relation between the GW speed and the graviton mass, and (b) Graviton mass normalized posterior. Horizontal/vertical lines and bands show the mean and 68\% confidence intervals, respectively: Tensor (red solid, `/' hatches), Vector (blue dashed, `\textbackslash' hatches).}
\label{fig:mg}
\end{figure}

Furthermore, as teased in Section \ref{sec:subluminal_gws}, GW speed constraints translate to a graviton mass bound. Working with this relation, we find the sampled graviton mass posteriors using the PTA correlations. The results are shown in figure \ref{fig:mg}.

This makes clear the relation between the GW speed and the graviton mass where now the PTA constraint on the GW speed translates to a graviton mass posterior: $m_{\rm g} \simeq \left(1.3105 \times10^{-22}{\rm \ eV}\right) \times f\left[{\rm yr}^{-1}\right] \times \sqrt{1 - v^2}$. The peak in the graviton mass at $m_{\rm g}\sim1.3 \times10^{-22}{\rm \ eV}$ thus relates to $v\sim 0.1$ where we see the GW speed samples cluster around. This similarly explains the squeezed shape of the graviton mass posterior as the slim mass range $m_{\rm g} \sim (1.3-1.31) \times 10^{-22}{\rm \ eV}$ maps with $0 < v \lesssim 0.1$ whereas the rest of $m_{\rm g} \lesssim 1.3 \times 10^{-22} {\rm \ eV}$ goes to $0.1 \lesssim v \leq 1$. It is also worth stressing that this relation is nonlinear, and so the shapes of the posteriors do not transform trivially. This explains why the mass upper limit $m_{\rm g} \sim 1.4 \times 10^{-22}$ eV, where the probability is nearly vanishing, do not manifest as an equivalently nearly vanishing GW speed probability at $v \ll 1$. We also note that the smallest $v$ considered is $v = 0.01$, which converts to a mass upper bound $m_{\rm g} \simeq 1.31\times 10^{-22}$ eV during the sampling. A related subtle point is that as far as the sampling on $m_{\rm g}$ goes $m_{\rm g} \lesssim 1.3105 \times 10^{-22}{\rm \ eV}$, where the upper bound is determined by the GW frequency that PTAs are sensitive to ($f \sim 1 {\rm \ yr}^{-1}$). Values that seemingly lie outside this in figure \ref{fig:mg} are only an artefact of the kernel density estimator considered to visualize the sampled posteriors \cite{Torrado:2020dgo}.

A subluminal GW speed in the PTA GW band has important implications for fundamental physics. Together with the LIGO/Virgo constraint \cite{LIGOScientific:2017zic}, this gives rise to dispersive nature of GWs, which could be associated with massive gravity or scalar-tensor theories with kinetic couplings, e.g., $S[g, \phi] \sim \left(\partial \phi\right)^2 R$, in the gravity sector. This revives interest in the couplings that were let go in favor of trivially satisfying the LIGO/Virgo constraint at the subkilohertz band \cite{Ezquiaga:2017ekz, Baker:2017hug}. Since nontrivial couplings also come with predictions in other regimes, such as cosmology \cite{Kase:2018aps, Ferreira:2019xrr} or strong fields \cite{Tattersall:2018map, Minamitsuji:2022vbi}, this would tie in PTA observations with other experiments to make stronger physics constraints.

\subsection{Subluminal GWs vs HD vs GW monopole}
\label{subsec:subluminal_hd_mon}

We take a look at subluminal GWs within the broader context of the Hellings-Downs (HD) curve and the GW-like monopole, or for short GW monopole \cite{NANOGrav:2021ini}. These give rise to spatial correlations with well established analytical forms,
\begin{equation}
\label{eq:Gab_hd}
    \Gamma_{ab}\left(\zeta\right) = \dfrac{3}{2} \left[ \dfrac{1}{3} + \left( \dfrac{1 - \cos \zeta}{2} \right) \left( \ln \left( \dfrac{1 - \cos \zeta}{2} \right) - \dfrac{1}{6} \right) \right]
\end{equation}
for the HD correlation, and
\begin{equation}
\label{eq:Gab_gwmon}
    \Gamma_{ab}\left(\zeta\right) = \dfrac{\delta_{ab}}{2} + \dfrac{1}{2} \,,
\end{equation}
for the GW monopole. The $\delta_{ab}$ in \eqref{eq:Gab_gwmon} is considered to take care of the small scale power that is dropped when pulling the pulsars to infinity. In \cite{Bernardo:2022vlj}, we have proposed a physical mechanism for the GW monopole by means of a scalar field \cite{Appleby:2022bxp}. Similarly, for the HD, $\Gamma_{aa} = 1$ is set in by hand to account for the small scale power \cite{NANOGrav:2021ini, Ng:2021waj}.

These traditional correlation signals were given their due consideration by PTAs \cite{NANOGrav:2021ini} and so it is important to assess the statistical significance of subluminal GWs with respect to these correlations. We model their cross correlated power in the PTA by multiplying the relevant ORFs, $\Gamma_{ab}\left(\zeta\right)$, by the GW amplitude squared, $A^2$. In contrast with subluminal tensors, both HD and GW monopole carry one parameter, $A^2$, that needs to be fit with the correlation observations.

We assess the statistical significance of subluminal tensors, HD, and the GW monopole by means of the reduced chi--squared statistic, $\overline{\chi}$, often considered the baseline for quantifying models' goodness of fit. We also consider the Akaike information criterion (AIC) and the Bayesian information criterion (BIC) \cite{Trotta:2008qt, Liddle:2007fy}, which in addition to providing a measure of the fit, also penalizes unnecessary addition of free parameters. To put briefly, the smaller $\overline{\chi}$, AIC, and BIC are, the better the fit. The relevant statistics of the sampling for the subluminal GW modes as well as for the HD and the GW monopole are summarized in Table \ref{tab:margstats}.

\begin{table}[h!]
    \centering
    \caption{Marginalized statistics for subluminal GW cross correlations constrained by PTA \cite{NANOGrav:2020bcs}. Results for the HD and the GW-like monopole ({GW} mon) are presented for comparison. The performance statistics (chi-squared, AIC, and BIC \cite{Trotta:2008qt, Liddle:2007fy}) are relative to the {GW} monopole, i.e., a positive value means statistical preference over the {GW} monopole.}
    \begin{tabular}{|r|r|r|r|r|r|} \hline
    \phantom{$\dfrac{1}{1}$} mode \phantom{$\dfrac{1}{1}$} & \phantom{111} $v$ \phantom{111} & $A^2$ [$\times 10^{-30}$] & $\Delta \overline{\chi}^2$ & $\Delta$AIC & $\Delta$BIC \\ \hline \hline 
    \phantom{$\dfrac{1}{1}$} Tensor \phantom{$\dfrac{1}{1}$} & $0.45^{+0.17}_{-0.42}$ & $7.4^{+2.1}_{-2.4}$ & $0.06$ & $-1.01$ & $-1.79$ \\ \hline 
    \phantom{$\dfrac{1}{1}$} Vector \phantom{$\dfrac{1}{1}$} & $0.39^{+0.14}_{-0.37}$ & $9.2^{+2.4}_{-6.0}$ & $-0.11$ & $-3.76$ & $-4.53$ \\ \hline 
    \phantom{$\dfrac{1}{1}$} HD \phantom{$\dfrac{1}{1}$}& $v = 1$ & $3.9 \pm 1.1$ & $-0.10$ & $-1.66$ & $-1.66$ \\ \hline 
    \phantom{$\dfrac{1}{1}$} {GW} mon \phantom{$\dfrac{1}{1}$} & ------- & $2.0 \pm 0.5$ & $0$ & $0$ & $0$ \\ \hline 
    \end{tabular}
    \label{tab:margstats}
\end{table}

We referred to the GW monopole since this phenomenological model has a competitive S/N in the NANOGrav data \cite{NANOGrav:2021ini}. More concretely, we expressed the metrics ${\rm X} \left(= \chi^2, {\rm AIC}, {\rm BIC}\right)$ in Table \ref{tab:margstats} as $\Delta {\rm X} = {\rm X}_{\rm GW mon} - {\rm X}_{i}$ such that if ${\rm X}_i < {\rm X}_{\rm GW mon}$, or that correlations model $i$ fits better compared with the GW monopole, then $\Delta {\rm X} > 0$, and vice versa. In short, a value in the more positive is better. At a glance, Table \ref{tab:margstats} shows almost all $\Delta \overline{\chi}^2$, $\Delta {\rm AIC}$, and $\Delta {\rm BIC}$ to be negative, meaning that none of the tensor and vector correlation models fit the data better than the GW monopole.

The GW monopole correlations represent the current data quite well, in agreement with earlier result \cite{Bernardo:2022vlj}. We emphasize that the GW monopole is physically distinct from the systematic monopole as it brings small scale power, and can be provided by a physical mechanism \cite{NANOGrav:2021ini, Bernardo:2022vlj}. In short this is a clear cut from the clock errors that were disfavored by the PTAs \cite{NANOGrav:2020bcs, Goncharov:2021oub}.

Letting the tensors move off the light cone improve the likelihood of their correlations in the data; however, this comes at a price since when considering the AIC and BIC, tensor correlations turn out disfavorable over the GW monopole. This means that there is no evidence for tensor correlations, may it be subluminal ones or the HD. The error bars are undesirably big to support or falsify tensorial spatial correlations. On the other hand, the results show that vector correlations are disfavored, in all significance metrics we consider. This improves the overall constraints on the space of vector correlations in PTA \cite{NANOGrav:2021ini}.

\begin{figure}[h!]
    \centering \includegraphics[width = 0.475 \textwidth]{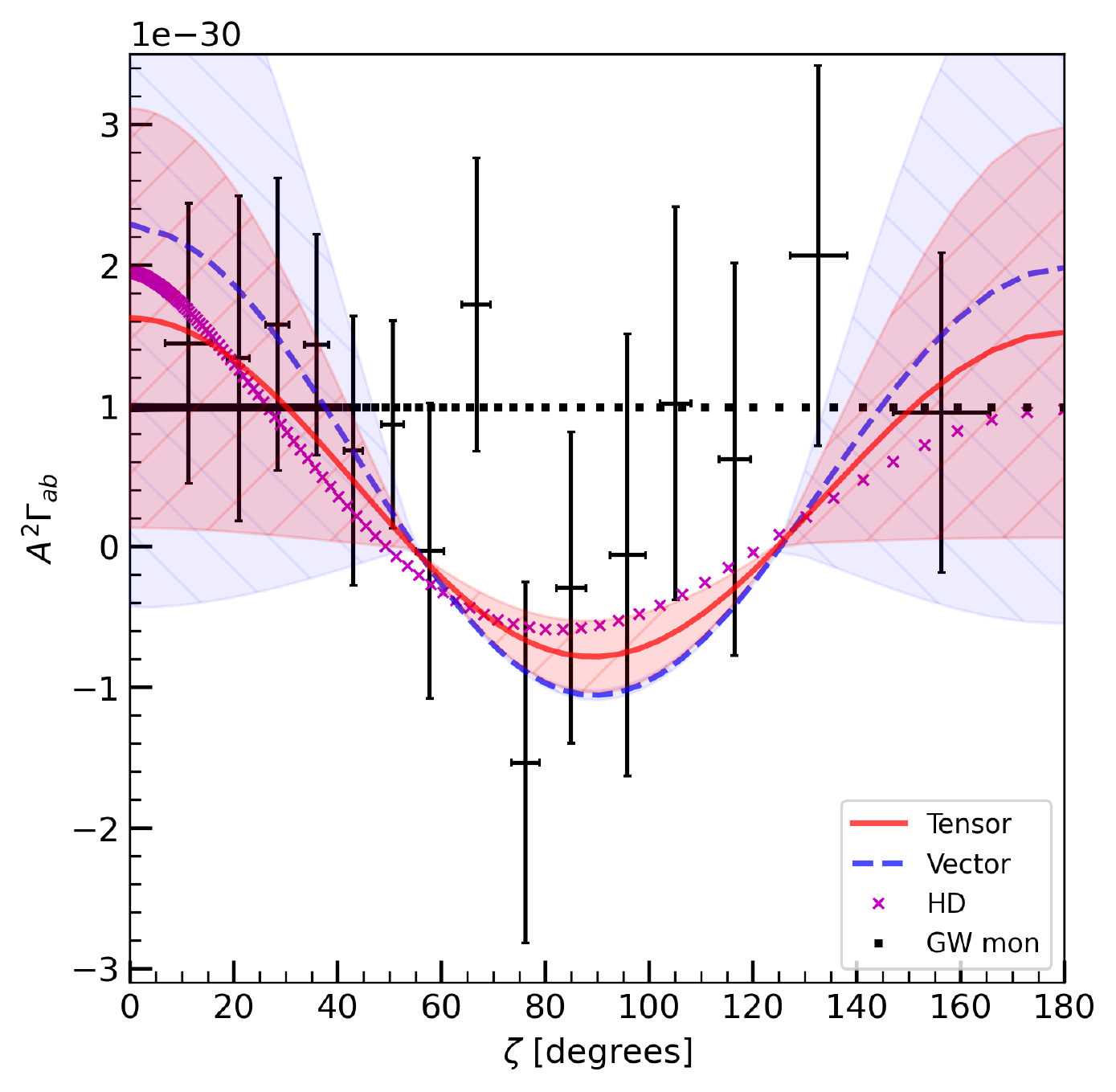}
    \caption{The ORF and its cosmic variance of the best fit subluminal GW modes: Tensor (red solid, `/' hatches), Vector (blue dashed, `\textbackslash' hatches). The hatched color bands show $1-\sigma$ cosmic variance uncertainty from the mean of the sampled posterior. The cross and square markers show the corresponding HD and GW monopole best fit correlations determined by sampling the same data points.}
    \label{fig:bestfits}
\end{figure}

We visualize the mean correlation curves that resulted from the sampling together with the data in figure \ref{fig:bestfits}.

Subluminal vector correlations in the SGWB are shaped at large angles by its dominant multipoles \cite{Bernardo:2022rif}
\begin{equation}
    C_2({\rm quadrupole}) \gtrsim C_1({\rm dipole}), C_3({\rm octupole}) > \cdots \,,
\end{equation}
an order that follows from the formulas in Section \ref{sec:pta_correlations}A. Near luminal GW speeds for vector correlations are also more strongly disfavored than tensor ones, as the vector correlations diverge at $\zeta \rightarrow 0^+$ \cite{Bernardo:2022rif, Bernardo:2022vlj}. Figure \ref{fig:bestfits} shows that the extra dipolar power, $C_1 \neq 0$, in the vector modes does not help out in making their correlations a better fit to the data, or at least not better than its tensor counterpart ($C_1 = 0$ and $C_2 > C_3 > \cdots$). Of course, we expect vector GW correlations to be irrelevant, for instance, through the arguments raised in \cite{Wu:2023pbt} that there are not enough mechanisms that lead to their excitation. We also add that vector modes are diluted by cosmic expansion. On the other hand, our result disfavors the vector correlations in the SGWB not by argument but by comparing directly with data.

The large uncertainty in the correlations clearly benefit the GW monopole, which so far has the largest signal-to-noise ratio among various models \cite{NANOGrav:2021ini}. Nonetheless, the tensor correlations should eventually manifest in the data as more pulsars are included \cite{Pol:2022sjn}. The cross correlations enhance with the square of the number of pulsars in a PTA, or, in other words, more pulsars guarantee smaller uncertainties in the cross correlations.

Upper bounds on the graviton mass can be obtained through the PTA correlations regardless of the significance of tensor modes. Particularly this work puts a bound of $m_{\rm g} \lesssim 10^{-22}$ eV on the graviton mass. This is consistent with an earlier result \cite{Wu:2023pbt} restricted within the framework of massive gravity \cite{Liang:2021bct}. Here, they evaluate the significance of tensor correlations per fixed graviton mass, but find that the values fall short to make any strong conclusions in support of massive gravity correlations. On the other hand, this work is conservative to the physics that endows the graviton a nonzero mass \cite{Montani:2018iqd, Poddar:2021yjd}, or rather follows a more framework--liberated direction in the sense that there is no gravity model in particular that is being considered. We examined GW propagation in this way, considering both tensor and vector correlations in the SGWB, by constraining the GW speed and the overall SGWB amplitude directly through the correlation measurements. It should be mentioned that the dispersion relation that links this work to massive gravity can also arise in various other models particularly in the weak field limit relevant for PTAs. In any case, neither present evidence in support of massive gravity or subluminal GWs in the data, and so are only able to give upper bounds to the graviton mass cast by the PTA GW energy scale.

\section{Conclusions}
\label{sec:conclusions}

We have explored subluminal GW modes as a possible source of spatial correlations in a PTA. This adds to the building literature on their feasibility given the available data: \cite{NANOGrav:2021ini} studied luminal scalar, vector, and tensor modes; \cite{Chen:2021wdo} on luminal scalar transverse modes; \cite{Bernardo:2022vlj} on subluminal scalar transverse and scalar longidutinal modes; \cite{Wu:2023pbt} on subluminal tensors. This work considered subluminal tensor and, for the first time, to the best of our knowledge, subluminal vector modes that may anchor as SGWB correlations \textit{within} the scope of the data.

Tensor correlations in general turned out to have low significance, which is understandable given the large uncertainties in the correlations data. Vector correlations appeared strongly disfavored, if not ruled out, by the data. Of course the result with vectors may not be surprising as we expect such modes to be diluted by the cosmic expansion, and that there's not too much known mechanisms that lead to their excitation \cite{Wu:2023pbt}. Our result on the other hand disfavors vector GW modes directly from the data without relying on other assumptions.

We emphasize that we use here the actual correlation observations to constrain the properties of the SGWB, rather than through the spectral density. The main advantage of which is that the SGWB produces a very distinct correlations signature that cannot be mistaken for systematics. We see this way also forecasts PTA/SGWB science's precision era as the cross correlations significantly enhance in the data with more pulsars \cite{Pol:2022sjn}.

We acknowledge that this analysis, and all that came before, are based on the hypothesis that the correlations in PTA are \textit{mainly} due to GWs. This is reasonable as GWs are real and thus by their wavelike nature it should not come as a surprise to find them in superposition, as in the SGWB. The challenge comes from setting apart the uncorrelated common process, $A_{\rm CP}$, that is confirmed by all PTAs, from the correlations, $\Gamma_{ab}$, which we associate with GWs, since PTA correlation measurements come in the combination $A^{2}_{\rm CP} \Gamma_{ab}$. We are hopeful that the observation of more pulsars \cite{Pol:2022sjn} and the recent IPTA checklist \cite{Allen:2023kib} should be able to settle the reservations surrounding this fundamental assumption.

Subluminal GWs open up various prospects. If the evidence for them improves in future data sets, then fundamental questions about the microscopic and cosmological theory of gravity naturally follow. Alternatively, should subluminal GW propagation be ruled out in PTA, then this may strongly put to rest further questions about the GW speed, since the PTA GW band and energy scale are orders of magnitude away from LIGO/Virgo characteristic scales. Tangential to this, the sources of the SGWB, which exist at very high energies, remain to be given further study once there is enough resolution in the data. On a practical side, incorporating the theoretical uncertainties in the correlations \cite{Allen:2022dzg, Allen:2022ksj, Bernardo:2022xzl} in the SGWB search pipelines of PTAs remain for future work. 

\medskip

\textit{Note Added.} \cite{Wu:2023pbt} appeared in the arXiv while we were writing ours.

\acknowledgments

This work was supported in part by the National Science and Technology Council (NSTC) of Taiwan, Republic of China, under Grant No. MOST 111-2112-M-001-065.


%

\end{document}